# Identifying Differences In Diagnostic Skills Between Physics Students: Developing A Rubric

A. Mason[1], E. Cohen[2], E. Yerushalmi[2], and C. Singh[1]

[1]Department of Physics and Astronomy, University of Pittsburgh, Pittsburgh, PA 15213, USA
[2]Department of Science Teaching, Weizmann Institute of Science, Rehovot, Israel

**Abstract.** Expert problem solvers are characterized by continuous evaluation of their progress towards a solution. One characteristic of expertise is self-diagnosis directed towards elaboration of the solvers' conceptual understanding, knowledge organization or strategic approach. "Self-diagnosis tasks" aim at fostering diagnostic behavior by explicitly requiring students to present diagnosis as part of the activity of reviewing their problem solutions. We have been investigating how introductory physics students perform in such tasks. Developing a robust rubric is essential for objective evaluation of students' self-diagnosis skills. We discuss the development of a grading rubric that takes into account introductory physics students' content knowledge as well as analysis, planning and presentation skills. Using this rubric, we have found the inter-rater reliability to be better than 80%. The rubric can easily be adapted to other problems, as will be discussed in a companion paper.

## INTRODUCTION

Bereiter & Scardamalia (1989) [1] argue in favor of intentional learning, namely a cognitive activity that has learning as a goal. Accordingly, instruction should make sure that students will not only focus their attention on the task, but also on learning from it.

In the context of a problem solving process, solvers use self-monitoring questions to elaborate the solution between successive trials. Yet, while self-monitoring is directed mainly towards arriving at a solution, it might also involve self-diagnosis directed towards more general learning goals such as elaboration of the solver's conceptual understanding [2].

We report the analysis of data obtained in a research study focused on self-diagnosis in the context of an abundant activity in physics problem solving: reviewing the solution that the learner has composed in order to improve it or learn from it. The pertinent questions are: 1) what learning outcomes result from this activity, and 2) how can instruction enhance the learning outcomes?

We constructed an alternative-assessment task in which students are required to present a diagnosis (namely, identifying where they went wrong, and explaining the nature of the mistakes) as part of the activity of reviewing their quiz solutions. We shall call these tasks "self-diagnosis tasks".

Analysis of data requires a robust method of grading that would allow assessing students' solutions as well as students' self-diagnosis. In particular, one would like to assess the solvers' conceptual understanding as it is reflected in their self-diagnosis.

To this end, there are two approaches [3]. The 1st approach maps the student statements to a representation of an expert "ideal" knowledge representation, i.e., what correct ideas needed to solve the problem are reflected in student's solution and diagnosis. The 2nd approach attempts to describe the novice knowledge per se, i.e., what ideas the student believes are needed to solve the problem are reflected in his/her solution and diagnosis.

The scoring rubric developed was aligned with both approaches with the intent of using it to help score student subjects in the continuing study on self-diagnosis [4,5]. The rubric was designed to comply with common standards of objectivity and reproducibility, i.e., validity as determined by four experts in physics education who perceive it as measuring an appropriate performance of the solution and self diagnosis and a high inter-rater reliability.

This paper details the final version of this rubric and its potential use. A companion paper will describe an analysis of student self-diagnosis for which the rubric was used, thus providing more detail.

## INITIAL CONSIDERATIONS

### Expert "ideal" knowledge representation

There were two main factors we desired to measure with this rubric. The first consideration is the student's use and application of physical principles. The student must invoke appropriate principles as well as apply them correctly in order to solve the problem. The second consideration is the student's presentation of a strategic problem-solving approach [6]. We are interested in evaluating if the student presented a helpful description of the problem's situation in terms of physics concepts and principles, e.g., if a diagram is drawn to help visualize the problem. In addition, it is of interest to see if the student constructed a good plan for solving the problem with regard to the target quantity and intermediate problem steps needed to obtain this quantity. Finally, we would like to evaluate if the student checked the reasonableness of his or her answer once it is obtained so as to make sure he or she did the problem correctly.

These considerations are generic, meaning that a rubric may be designed with a set of general guidelines in place. However, it should also be possible for specific attributes of the problem to be added to or removed from the rubric as befits the problem.

## Sample Problem Used

The problem used features a girl of mass $m_{girl}$ riding a rollercoaster that consists of a steep hill of height $h_0$ followed by a circularly shaped bump with height $h_f$ and reflecting a circular radius $r$, and asks, given that the girl is sitting on a scale on the rollercoaster cart, how much will the scale read at the top of the circular bump?

In order to solve this problem, a student will need to understand that the target variable is the normal force the scale exerts on the girl. To calculate the normal force at point B, the student will have to invoke Newton's 2$^{nd}$ Law: $\Sigma \vec{F} = m \cdot \vec{a}$. He will need to find as intermediate variables the net force (sum of the force of gravity and the normal force) and the acceleration at this point. To calculate the acceleration, he will have to invoke the expression for centripetal acceleration: $a_g = v_B^2/R$. The intermediate variable, the speed of the cart at the top of the circular bump, can be found using the law of conservation of mechanical energy, $PE_i + KE_i = PE_f + KE_f$, between point of departure and the top of the bump (which is justified because all forces doing work are conservative forces). A complete description of the problem and its solution is available in Ref. [4].

Figures 1A and 1B respectively represent the quiz and self-diagnosis attempt of a student in the self-diagnosis studies. This student's work features examples of what a student might do correctly as well as some good examples of what he might have done incorrectly, which the rubric should be able to reflect.

## THE RESULTING RUBRIC

### Specific Knowledge: Categories and Subcategories

Table 1 represents the rubric that was ultimately developed. The three columns for each student represent the three ways in which the student's work is evaluated for a self-diagnosis experiment: from left to right, they represent the researcher diagnosis of the student quiz solution (RDS), the student's self-diagnosis of his/her solution (SDS), and the researcher's judgment of this student's self-diagnosis (RSD). For each column, the students are evaluated by a series of criteria represented by the rubric's rows. The rows of the rubric are divided into three main categories: physical principles (hereafter referred to as "physics" for the sake of brevity), problem solving presentation (referred to as "presentation"), and algebra ("math").

**FIGURE 1A.** Sample student B14's quiz from the self-diagnosis study. The circled numbers 1 and 2 are references from the student's self-diagnosis as labeled in figure 1B, and the circled number 14 is the code number for the student.

**FIGURE 1B.** Sample student B14's self-diagnosis from the self-diagnosis study. The other groups did not receive this worksheet but instead wrote their diagnosis elsewhere.

**TABLE 1.** Rubric developed for self-diagnosis study. The student featured in Figures 1A and 1B is graded to serve as an example. Abbreviations: RDS = Researcher diagnosis of solution; SDS = Student's diagnosis of solution; RSD = Researcher judgment of student's diagnosis. In the RDS/RSD "+" is given if a student correctly performs/identifies a mistake defined by some subcategory. A "-" is given if the student incorrectly performs or fails to identify a mistake or identify it incorrectly. If a student is judged to have gotten something partially correct, then the grader may assign ++/-, +/-, or +/--. The term "n/a" is assigned if the student could not reasonably address a subcategory given the prior work done. For example, the sample student correctly invoked conservation of energy on the original quiz and therefore did not address it during self-diagnosis. In the researcher's judgment column, the grader would then state "n/a" and not consider this invoked law in assigning a grade for the researcher's judgment. In the SDS column + and – reflect how the students perceive themselves to be correct/wrong on the quiz solution. An "x" indicates the student did not address a subcategory at all, which is interpreted as the students perceiving themselves to be correct on the quiz solution.

| General Task | Specific Criteria | | RDS | SDS | RSD |
|---|---|---|---|---|---|
| **PHYSICS PRINCIPLES (Ph.)** | | | | | |
| | **Ideal knowledge** | | | | |
| **Invoking physical principles** | 1. Conservation of mechanical energy (student correct on quiz) | | + | x | n/a |
| | 2. Non-equilibrium applications of Newton's second law in centripetal motion | | - | - | + |
| | 3. Justification for CE (e.g. all forces doing work are conservative since non-conservative forces are perpendicular to the path of motion) | | - | x | - |
| | **Novice knowledge per se** | | | | |
| | 4. inappropriate principle: "-" marked if inappropriate principle is used in student's solution or diagnosis (student used gravitational law here) | | - | - | + |
| **Applying physical principles** | **Ideal knowledge** | **Novice knowledge per se** | | | |
| | | defining the system inappropriately or inconsistently | - | - | + |
| | 1. conservation of mechanical energy | e.g. calculation of KE/PE without energy conservation | - | - | + |
| | 2. Non-equilibrium applications of Newton's second law | e.g. referring to centripetal force as a physical force | | | |
| | | e.g. forgetting normal force | (-) | - | + |
| | | (etc.) | | | |
| **ALGEBRA (Alg.)** | | | | | |
| Algebra | Algebraic manipulation | | + | x | n/a |
| **PRESENTATION (Pre.)** | | | | | |
| **Description (Des.):** | 1. Invokes a visual representation | FBD, acceleration vector, axis, defining PE = 0, radius of the circle | -, -, -, +, + | -. x, x, x, x | +, -, -, n/a, n/a |
| | 2. Clear/appropriate knowns (the student listed most but missed some) | | ++/- | + | ++/- |
| **Plan/Solution Construction (Plan):** representing the problem as a set of sub-problems | 1. Appropriate target quantity chosen (first chose $F_g$, fixed in diagnosis) | | - | - | + |
| | 2. did not write down surplus equations or intermediate variables | | - | x | - |
| | 3. appropriate intermediate variables explicitly stated ($v_b$ but not $a_c$) | | +/- | +/- | + |
| | 4. explicitly stating in words or a generic form the principles used to solve for this intermediate variables (not done by student) | | - | x | - |
| **Evaluation (Che.)** | 1. writing down the units (student sometimes did so) | | +/- | x | +/- |
| | 2. checking the answer | | - | - | + |

The physics category is divided into two subcategories: invoking a physical principle and applying that principle. Each row in each subcategory therefore represents every physical principle that a student will have to invoke and apply to correctly solve the problem. For example, in the problem described here for which the introductory students are evaluated, conservation of energy and a non-equilibrium application of Newton's 2$^{nd}$ Law in circular motion are both required. Therefore, there will be two rows in the invoking subcategory to evaluate if the student cited these laws, and two corresponding rows in the applying subcategory to evaluate if the student applied them correctly. We unified Newton's 2$^{nd}$ Law and centripetal acceleration as one subcategory, as those were difficult to differentiate in students' answers. In the applied principle section, we consider how well the student does on grading according to both expert and novice representations by noting what specific errors the student makes and evaluating the student's diagnosis of the errors.

Also included in the physics category are two other kinds of criteria. The first deals with any justifications that a student should cite for invoking or applying a physical principle. This is determined a priori according to the needs of each problem. In the given case, students are expected to justify invoking conservation of energy in the aforementioned problem as follows: energy is conserved as the only forces doing work on the girl "system" are conservative forces (gravitation). The normal (non-conservative) force is perpendicular to the cart's motion, and hence does no work. There is an additional row in the invoked physics subsection that tracks if the student invoked an inappropriate principle that doesn't apply to the problem, whether legitimate or invalid.

The plan/solution category has three different subcategories. Problem description involves anything the student may do to facilitate properly understanding the problem question. This includes drawing visual representations and also listing all given known quantities clearly and appropriately. Planning and solution construction tracks the student's steps, and checks to see if the student has done the following: described the appropriate target variable for which the student is solving, avoided writing down surplus equations, described appropriate intermediate variables to reach the target variable, and explicitly stated some methodology used to take the steps used in the problem. Evaluation involves the student's tracking of his or her own work, usually involving a check of the answer and taking care to write down the proper units. Note that the presentation and algebra subcategories are essentially general and do not have to be changed from problem to problem; only the specific criteria based upon the subcategories need to be changed.

The algebra category contains one row that is included to check for mathematical mistakes made during the problem-solving process, for example forgetting a coefficient when rewriting an equation. This row shows whether a student makes a minor math error, or if a student attempts algebraic manipulation to attain a desired quantity.

## Scoring

The scoring reflects researcher priorities. The 1st approach mentioned in the introduction, in which a researcher assess how student's knowledge compares to the ideal knowledge, would possibly weigh each ideal subcategory as worth 1 point if "+" and worth 0 points if "-". The 2nd approach, in which a researcher assesses in what detail students are able to identify their mistakes, would possibly also weigh each novice subcategory. Initially, we took the 1st approach when we scored the quiz solution and the self-diagnosis (RSD). Later on, we took the 2nd approach for the RSD as it allowed better differentiating between students.

Table 2 displays how the marks given in Table 1 are interpreted as an overall score. The overall grade for a physics score can be interpreted as an average of all possible criteria that the student correctly addressed. Eventually, we did not score the "justification" part (row 4), since students' justification of chosen principles was found to occur very rarely. Therefore, we assume that the students did not think of justification as part of the solution procedure, and by extension, the self-diagnosis procedure.

| TABLE 2. Rubric scoring for the sample student. | |
| --- | --- |
| **Category** | **Grading** |
| RDS | **Ph: 0.17; Pre: 0.27** |
| SDS | **Ph: 0.14; Pre: 0.65** |
| RSD | **Ph: 1; Pre: 0.54** |

## Reliability

Part of the purpose of this rubric is to assure objectivity in grading. With this in mind, two researchers independently graded ~10 sample students using the rubric for the problem diagrammed in Figures 1A and 1B. They then discussed how criteria should be applied to the students' work objectively. We found that the graders could agree to within at least 80% of each other in grading the rubric. This established a reasonable inter-rater reliability that was consistent in a more thorough analysis of about 200 students. The companion paper will outline the use of this rubric to examine the effect of self-diagnosis of a quiz performance on future exams.

## ACKNOWLEDGMENTS

The research for this study was supported by ISF 1283/05 and NSF DUE-0442087.